\documentstyle[aps,prl,multicol,epsf]{revtex}
\draft
\begin{document}              
\title{Mesoscopic fluctuations of the Coulomb drag}
\author{B. N. Narozhny and I. L. Aleiner}
\address{Department of Physics and Astronomy,SUNY Stony 
Brook, Stony Brook, NY 11794}
\address{Theoretische Physik III, Ruhr-Universit\"at 
Bochum, 44780 Bochum, Germany}
\maketitle

\begin{abstract}
We consider mesoscopic fluctuations of the Coulomb drag 
coefficient $\rho_D$ in the system of two separated 
two-dimensional electron gases. It is shown that at low 
temperatures sample to sample fluctuations of $\rho_D$ 
exceed its ensemble average. It means that in such a regime
the sign of $\rho_D$ is random and the temperature 
dependence almost saturates $\rho_D \sim 1/\sqrt{T}$.
\end{abstract}

\date{Draft: \today}
\begin{multicols}{2}

When two electronic layers are brought close to each other 
to form a bi-layer system, a current flowing through one 
of the layers(the active layer) is known to induce a 
voltage $V_D=\rho_D I$ in the other (passive) layer 
\cite{the1,the2,exp1,exp2}. The effect, which is called 
the drag, was first predicted theoretically 
\cite{the1,the2} in the model where the carriers in the two
spatially separated layers interacted via long-range 
Coulomb interaction. Experimentally the Coulomb drag was 
first observed in a three-dimensional electron gas layer 
while the current was driven through a two-dimensional 
electron gas (2DEG) \cite{exp1}. Subsequent experiments 
studied the effect in 2DEG bi-layers \cite{exp2}, 
electron-hole \cite{exp3}, and 
normal-metal-superconductor systems \cite{exp4}. More 
recently, the drag was studied in the 2DEG bi-layer system 
in high magnetic field \cite{exp6}. 

The quantity, which is studied theoretically is the 
transconductance $\sigma_D$. To the lowest non-vanishing 
order in the interlayer interaction it is proportional to 
the drag coefficient $\rho_D$ ($\sigma_i$ is the Drude 
conductance of the $i$-th layer) 

\begin{eqnarray*}
\rho_D \approx {{\sigma_D}\over{\sigma_1\sigma_2}}.
\end{eqnarray*}

\noindent
As a function of temperature the observed $\sigma_D$ 
roughly follows the quadratic law $\sigma_D \sim T^2$, 
although the ratio $\sigma_D/T^2$ deviates from the 
constant value \cite{exp2,exp6}.

The $T^2$ dependence of the Coulomb drag coefficient 
follows from the Fermi liquid phase space argument. To 
create a current in the passive layer, it is necessary to 
create a pair of electron-like (filled states with energy 
greater than the Fermi energy $\epsilon > \epsilon_F$) and 
hole-like excitations (empty states $\epsilon <\epsilon_F$)
in a state with non-zero momentum. The energy and momentum 
of the pair come from an electron in the active layer, 
which is moving with the driving current. In each layer, 
the scattering states are limited to the energies of order 
$T$ relative to the Fermi level, which gives two powers of 
$T$ to the drag coefficient. However, the momentum is 
transferred equally to electrons and holes, therefore in 
the case of electron-hole symmetry the drag of the 
electrons cancels that of the holes. Thus the effect is 
non-zero only due to the electron-hole asymmetry. 
Similarly, the asymmetry is necessary for the electron and 
hole system in the active layer to have non-zero total 
momentum in the first place. The asymmetry can be expressed
as a derivative of the density of states $\nu$ and/or the 
diffusion constant $D$ with respect to the chemical 
potential $\mu$. This can be obtained rigorously in the 
diagrammatic formalism \cite{oreg}. For the case of 
diffusive layers the disorder-averaged transconductance is

\begin{equation}
\langle \sigma_D \rangle = {{e^2}\over{\hbar}} 
{{\pi^2}\over{3}}{{(\hbar T)^2}\over{g^2(\kappa d)^2}}
\left({{\partial}\over{\partial\mu}}
\left(\nu D\right)\right)^2\ln{{T_0}\over{2T}},
\label{av}
\end{equation}

\noindent
where for simplicity we take the layers to be identical, so
that they have the same chemical potential, diffusion 
constant and the dimensionless conductance 
$g=25.8 k\Omega /R_\Box $. The logarithm is cut at the 
scale $T_0 = D\kappa /d$ and $\kappa = 2\pi e^2 \nu$ is 
the inverse Thomas-Fermi screening length.

Such effects of the electron-hole asymmetry are well-known,
for instance the thermopower in disordered electronic 
systems \cite{tmp} or adiabatic pumping \cite{pmp}. As 
these effects are due to the electron-hole asymmetry, the 
average quantities are small, since each derivative with 
respect to the chemical potential brings one power of the 
Fermi energy $E_F$ to the denominator. On the other hand, 
the typical energy scale of mesoscopic effects is the 
Thouless energy $E_T = \hbar D/L^2$ ($L$ is the sample 
size), which is much smaller than the Fermi energy. 
Therefore the effects mentioned above exhibit mesoscopic 
fluctuations, much larger than the average.

In this Letter we show that the mesoscopic fluctuations of 
the Coulomb drag coefficient can indeed be larger than the 
average Eq.~(\ref{av}), even if the electron systems in 
both layers are good metals ($g\gg1$). To characterize the 
fluctuations, we calculate the average square of the 
(random) transconductance. The result of lengthy albeit
straightforward calculations shown below is given by

\begin{mathletters}
\begin{eqnarray} 
&&\langle\sigma_D^{\alpha\beta}
\sigma_D^{\alpha^\prime\beta^\prime}\rangle = 
(\delta^{\alpha\alpha^\prime}\delta^{\beta\beta^\prime}
+\delta^{\alpha\beta^\prime}\delta^{\alpha^\prime\beta}) 
\langle\sigma_D^2\rangle ,
\\
&&
\nonumber\\
&&
\langle\sigma_D^2\rangle = {{\gamma}\over{18\pi^3}} 
\left({{32\ln{2} - 14}\over{3}}\right)
{{e^4}\over{\hbar^2}}
{{E_T\tau_\varphi\ln{\kappa d}}\over{g^4(\kappa d)^3}},
\label{res2}
\end{eqnarray}
\label{res}
\end{mathletters}

\noindent
where the numerical factor $\gamma = 1.0086$ is the value 
of the integral

\begin{eqnarray*}
\gamma &&= {{1}\over{2}}\int\limits_0^{\infty} 
{{dx_1 dx_2 x_1 x_2}\over{(x_1^2 + x_2^2)}}
\left[J_0(x_1)J_0(x_2)+J_2(x_1)J_2(x_2)\right]
\nonumber\\
&&
\nonumber\\
&&
\left(K_0(x_1)K_0(x_2)+
\left[{2\over{x_1^2}}-K_2(x_1)\right]
\left[{2\over{x_2^2}}-K_2(x_2)\right]\right),
\end{eqnarray*}

\noindent
where $J_i(x)$ and $K_i(x)$ are the Bessel functions 
\cite{abr}. Here $\tau_\varphi \ll (E_T)^{-1}$ is the 
dephasing time. This result is valid in the most relevant 
regime $L_\varphi = \sqrt{D\tau_\varphi} \leq L$ and 
$\kappa d \gtrsim 1$. If $\kappa d \leq 1$, then the 
average square of the conductance is 
$\langle\sigma_D^2\rangle\propto{{e^4}\over{\hbar^2}}
{{E_T\tau_\varphi}\over{g^4}}$ with the coefficient of 
order unity. In what follows we first discuss the 
experimental consequences of our results, then explain it 
qualitatively and finally give the rigorous calculation. 
The fluctuations Eq.~(\ref{res}) depend on temperature 
only through the dephasing time 
$1/\tau_\varphi \approx T/g$ and at low enough temperatures
they should dominate the behavior of the transconductance. 
Therefore the $T^2$ decrease of $\sigma_D$ should at some 
small temperature $T_*$ be almost saturated at a 
sample-dependent value. Let us estimate $T_*$ for the 
samples used in existing experiments \cite{exp2,exp6} using
the reported parameters of the samples. Collecting the 
numerical factors, we write the ratio of the square of the 
average transconductance Eq.~(\ref{av}) and the averaged 
square Eq.~(\ref{res}) as

\begin{eqnarray*}
{{\langle\sigma_D\rangle^2}\over{\langle\sigma_D^2\rangle}}
= \left({{gT}\over{E_F}}\right)^4 
{{1}\over{E_T\tau_\varphi}}
{{20}\over{\kappa d \ln\kappa d}}.
\end{eqnarray*}

\noindent
We take the interlayer spacing to be $d=200\AA$ 
\cite{exp2,exp6}; the screening length in $GaAs$ is 
$\kappa^{-1} = 100\AA$; the Thouless energy is given by 
$E_T = g/(2\pi\nu L^2)$; and the dephasing time 
$\tau_\varphi^{-1} \simeq T\ln g /g$ \cite{aa}. Then we 
estimate $T_*$ as the temperature at which the ratio 
$\langle\sigma_D\rangle^2/\langle\sigma_D^2\rangle$ is 
equal to unity

\begin{eqnarray*}
T_* = E_F \left( 16\pi g^2 n L^2\right )^{-1/5} 
\approx 0.2 K,
\end{eqnarray*}

\noindent
where the Fermi energy in the samples \cite{exp2,exp6} 
$E_F \simeq 60 K$, the electron density 
$n=1.5\times 10^{11} cm^{-2}$, the size of the sample
$L \simeq 200 {\mu}m$, and the conductance is 
calculated from the sheet resistance of the sample 
$R=10\Omega /\Box$. 

The estimated $T_*$ is lower than the temperature range 
for the existing data \cite{exp2,exp6}, therefore there is 
no trace of the fluctuations Eq.~(\ref{res}) in the data. 
However, if one takes a dirtier sample, with the sheet 
resistance, for instance, $1k\Omega/\Box$, then the 
estimate for $T_*$ becomes $2K$ and the effect of the 
fluctuations becomes observable. To push $T_*$ even higher,
the sample size can be also reduced. 

Let us now explain Eq.~(\ref{res}) qualitatively. First, 
consider the lowest temperature regime $T \ll E_T$, so that
the sample is effectively zero-dimensional (0D). The
mesoscopic fluctuations of the usual conductance are 
universal $\delta\sigma \simeq e^2/\hbar$. The 
transconductance is associated with the interlayer 
interactions, thus possessing additional smallness. The value 
of such smallness can be estimated by the Golden rule
argument which is comprised by (i) phase volume; (ii) 
matrix elements; (iii) electron-hole asymmetry (dependence 
of the density of states on the energy). Matrix elements in
0D do not depend on energy and give smallness $1/g^2$ 
\cite{al}. Therefore the phase volume is limited by 
temperature only, which gives the factor $T^2$. Finally, 
the electron-hole asymmetry $\partial_\mu (\ln \nu(\mu))$ 
is the random quantity with the typical value $1/E_T$. 
Putting everything together, we arrive to the estimate

\begin{eqnarray}
{\mathrm r.m.s.} \delta\sigma_D \sim {{e^2}\over{\hbar}}
{{T^2}\over{(E_T g)^2}}
\label{est}
\end{eqnarray} 

{
\narrowtext
\begin{figure}[ht]
\vspace{0.2 cm}
\epsfxsize=6 cm
\centerline{\epsfbox{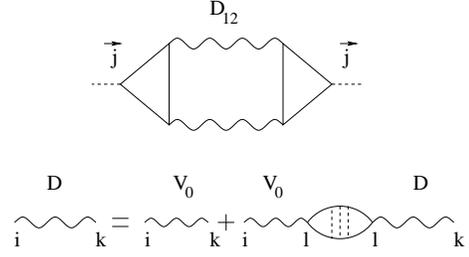}}
\vspace{0.5cm}
\caption{A. The diagram for the transconductance in the 
lowest non-vanishing order in the interlayer interaction. 
The solid lines are the exact Green's functions of the 
non-interacting electron system (in the presence of 
disorder) and the wavy lines are the disorder averaged 
interlayer Coulomb interaction propagators 
Eq.~(\protect\ref{int}). B. The RPA 
scheme for the calculation of the interaction propagators. 
The dashed lines denote the disorder. The indices 
$\protect i,k,l = 1,2$ indicate the layer.}
\label{1}
\end{figure}
}

At higher temperatures the averaging should be performed 
by dividing the sample into patches of the size 
$L_\varphi\times L_\varphi$, since on larger scales the 
phase coherence is destroyed. The contribution of each 
patch $\langle\delta\sigma_D^2(L_\varphi)\rangle$ is 
approximately the same and they can be simply comboned as 
a network of random resistors

\begin{eqnarray} 
\langle\delta\sigma_D^2\rangle = 
\langle\delta\sigma_D^2(L_\varphi)\rangle 
\left( {{L_\varphi}\over{L}} \right)^2
\label{net}
\end{eqnarray}

\noindent
Now, $\langle\delta\sigma_D^{\it rms}(L_\varphi)\rangle$ 
can be found similar to Eq.~(\ref{est}) with two important 
differences: (i) the fluctuations of the density of states 
are summed from the scale of order $T$, rather than 
$E_T(L_\varphi)$. 
This suppresses the fluctuations in each layer by the 
factor $\sqrt{E_T(L_\varphi)/T}$. (ii) The matrix elements 
become energy dependent on the energy scale larger than 
$E_T$, decreasing with the transmitted energy $\omega$ as 
$|M|^2 \sim 1/\omega^2$. As a result, the transmitted 
energy is limited by 
$\omega \sim E_T(L_\varphi) = 1/\tau_\varphi$, rather then 
by $T$, so in the estimate of the phase volume we should 
replace $T^2$ by $T/\tau_\varphi$. So, we find

\begin{eqnarray}
\delta\sigma_D(L_\varphi) \sim 
{{T \left({1\over{\tau_\varphi}}\right)}
\over{(E_T(L_\varphi)g)^2}} 
\left(\sqrt{{{E_T(L_\varphi)}\over{T}}}\right)^2 \sim 
g^{-2}
\label{patch}
\end{eqnarray}

\noindent
Finally, to estimate the total magnitude of the 
transconductance fluctuations we substitute 
Eq.~(\ref{patch}) into Eq.~(\ref{net}) to obtain

\begin{eqnarray*}
\langle \delta\sigma_D^2 \rangle \sim g^{-4} 
\left({{L_\varphi}\over{L}}\right)^2 \sim 
g^{-4}E_T \tau_\varphi.
\end{eqnarray*}

\noindent
This estimate yields the same result as Eq.~(\ref{res}) up 
to numerical factors and the dependence on $\kappa d$. 

Our results suggest the following picture of the Coulomb 
drag. If one starts measuring the drag coefficient at high 
$T$ and proceeds by lowering the temperature, then at 
first the transresistance will decrease roughly as $T^2$, 
as follows from Eq.~(\ref{av}). At the temperature $T_*$, 
estimated above, the transresistance will appear to 
saturate ($\sigma_D \propto 1/\sqrt{T}$), as the 
fluctuations Eq.~(\ref{res}) will start to dominate. The 
particular value of the prefactor will be sample dependent 
and, what is more important, will have random sign. If the 
temperature will be decreased further, then at very low 
temperatures $T<E_T$ the sample will effectively become 
zero dimensional and the $T^2$ decrease will be restored
(also with a random coefficient), so that at $T=0$ the 
drag coefficient vanishes.

Let us now present the calculation. The electrons in both 
layers interact via the Coulomb interaction. The 
interaction propagators corresponding to the dynamically 
screened Coulomb interaction can be obtained within the 
RPA scheme (see Fig. 1B) using the Green's functions of 
non-interacting electrons. For our purposes we only need 
the propagator of the interlayer interactions, which is 
given by (here we set the layers to be identical, so that 
they have the same density of states $\nu$ and diffusion 
coefficient $D$)

\begin{equation}
{\cal D}^R(\omega, Q) = {{1}\over{2\nu D Q^2}}
{{(-i\omega + D Q^2)^2}\over{-i\omega +(1+\kappa d)D Q^2}}.
\label{int}
\end{equation}

The transresistance in the disordered two-layer system can 
be expressedin terms of the exact Green's functions of 
non-interacting, disordered electron system and the 
interaction propagators Eq.~(\ref{int}). To the lowest 
non-vanishing order in the interlayer interaction the 
transresistance is given by the diagram Fig. 1A. The left
 and right triangles correspond to the two layers in the 
system and the wavy line is the interlayer interaction 
propagator Eq.~(\ref{int}). As the electron Greens's 
functions now depend on disorder, this $\sigma_D$ is a 
random quantity and its moments should be averaged over 
disorder. Before averaging, the expression for $\sigma_D$ 
corresponding to the diagram on Fig.~\ref{1} can be 
written as

\begin{equation}
\sigma_D^{\alpha\beta} = {{1}\over{4 V}} 
\int{{d\omega}\over{2\pi}}
\left({{\partial}\over{\partial\omega}} 
\coth{{{\omega}\over{2T}}}\right)
{\cal D}^R_{12}\Gamma^{\alpha}_{23}
{\cal D}^A_{34}\Gamma^{\beta}_{41},
\label{dia}
\end{equation}

\noindent
where the indices indicate the spatial coordinates. Points 
$1,2$ belong to one layer and $3,4$ to the other. The 
triangular vertices $\Gamma^{\alpha}$ are given by

\begin{mathletters}
\begin{equation}
\Gamma^{\alpha}_{12}(\omega) = 
\int{{d\epsilon}\over{2\pi}}
\left[J^{\alpha}_{12}(\omega,\epsilon) + 
J^{\alpha}_{21}(-\omega,\epsilon) + 
I^{\alpha}_{12}(\omega,\epsilon)\right],
\label{ing}
\end{equation}

\noindent
where

\begin{eqnarray}
J^{\alpha}_{12}&&(\omega,\epsilon) =
\left(\tanh{{{\epsilon - \omega}\over{2T}}} - 
\tanh{{{\epsilon}\over{2T}}}\right) 
\nonumber\\
&&
\nonumber\\
&&
\left[G^R_{12}(\epsilon - \omega) - 
G^A_{12}(\epsilon - \omega)\right]
\left[G^R(\epsilon) j^{\alpha} G^A(\epsilon) \right]_{21};
\end{eqnarray}

\begin{eqnarray}
I^{\alpha}_{12}(\omega,\epsilon) && =
\left(\tanh{{{\epsilon - \omega}\over{2T}}} - 
\tanh{{{\epsilon}\over{2T}}}\right) (r_1-r_2)^{\alpha}
\nonumber\\
&&
\nonumber\\
&&
\left[G^R_{12}(\epsilon)G^R_{21}(\epsilon - \omega) - 
G^A_{12}(\epsilon)G^A_{21}(\epsilon - \omega)\right].
\label{not}
\end{eqnarray}
\label{ver}
\end{mathletters}

\noindent
The exact electronic Green's functions used in 
Eq.~(\ref{ver}) can be written in terms of the exact 
wavefunctions of the system as

\begin{eqnarray*}
G^{R(A)}_{12}(\epsilon)=\sum\limits_j
{{\Psi^*_j(\vec{r}_1)\Psi_j(\vec{r}_2)}
\over{\epsilon-\epsilon_j\pm\imath 0}},
\end{eqnarray*}

\noindent
where $j$ labels the exact eigenstates of the system and
$\epsilon_j$ are the exact eigenvalues.

The known result for the averaged transconductance 
Eq.~(\ref{av}) is obtained \cite{oreg} by averaging the 
triangular vertices $\Gamma^{\alpha}$ independently for 
each layer (for the effect of the correlated disorder see 
Ref.~\cite{khve}). However, as we discussed above, in the 
intermediate temperature range the fluctuations exceed the 
average, and the temperature dependence saturates. To 
characterize the fluctuations we average the square of the 
transconductance. The mesoscopic fluctuations of the 
interaction propagators can be neglected, because they 
produce only the small fluctuating coefficient in 
Eq.~(\ref{dia}). Therefore, we have to average the product 
of two triangular vertices $\Gamma^{\alpha}$ (in the same 
layer; note that in this temperature regime Eq.~(\ref{not})
does not contribute). The corresponding diagrams are shown 
in Fig. 2. After averaging, each diagram contains six 
diffusons and one Hikami box (see Fig. 3). The calculation
of the arising 14-dimensional integral is greatly 
simplified by the following observation. The dominant 
contribution to the frequency integrals Eq.~(\ref{ing}) 
comes from the region, where the external frequencies 
[which are the frequencies of the interaction propagators 
Eq.~(\ref{int})] 
$\omega_1 + \omega_2 \simeq \kappa d /\tau_\varphi $ and 
$\omega_1 - \omega_2 \simeq 1 /\tau_\varphi$ so that the 
frequency difference is small compared to the sum. The 
momentum integral is dominated by the region 
$1/\tau_\varphi \le Q \le \kappa d/\tau_\varphi$. Since we 
are in the intermediate temperature regime $T > E_T$, the 
energy transfer $\omega$ is smaller than the temperature, 
and the vertices Eq.~(\ref{ver}) can be expanded in the 
inverse temperature. Now the dimensional analysis gives 
the resulting expression for the fluctuations 
Eq.~(\ref{res}), which depends on temperature only through 
$\tau_\varphi$ as we discussed above. The numerical 
coefficient can now be obtained by performing the 
integration without further approximations. The factor 
$\gamma$ comes from the angle integration and the 
numerical factor in Eq.~(\ref{res2}) from the integration 
over the small frequency difference.

{
\narrowtext
\begin{figure}[ht]
\vspace{0.2 cm}
\epsfxsize=7 cm
\centerline{\epsfbox{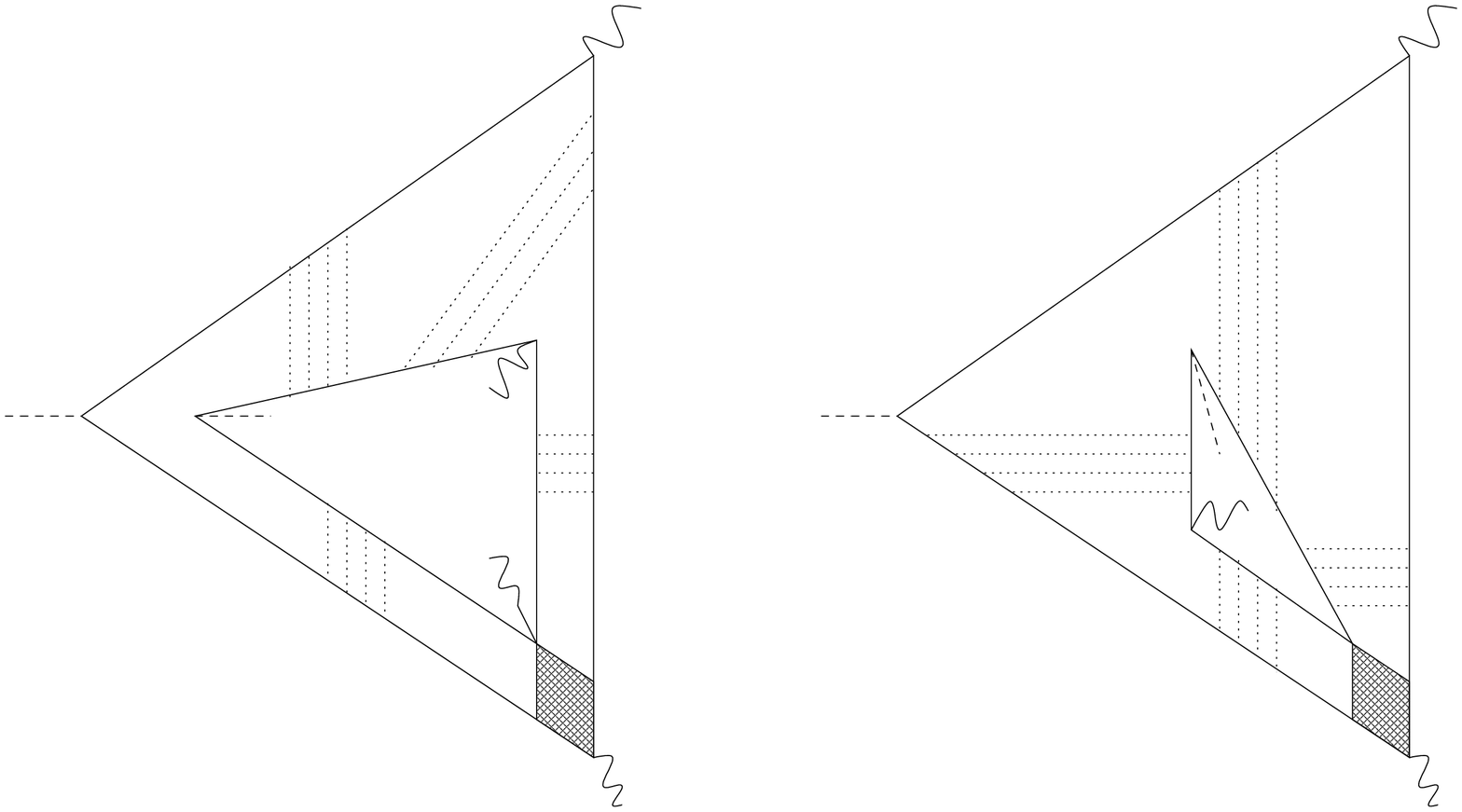}}
\vspace{0.2cm}
\epsfxsize=3.5 cm
\centerline{\epsfbox{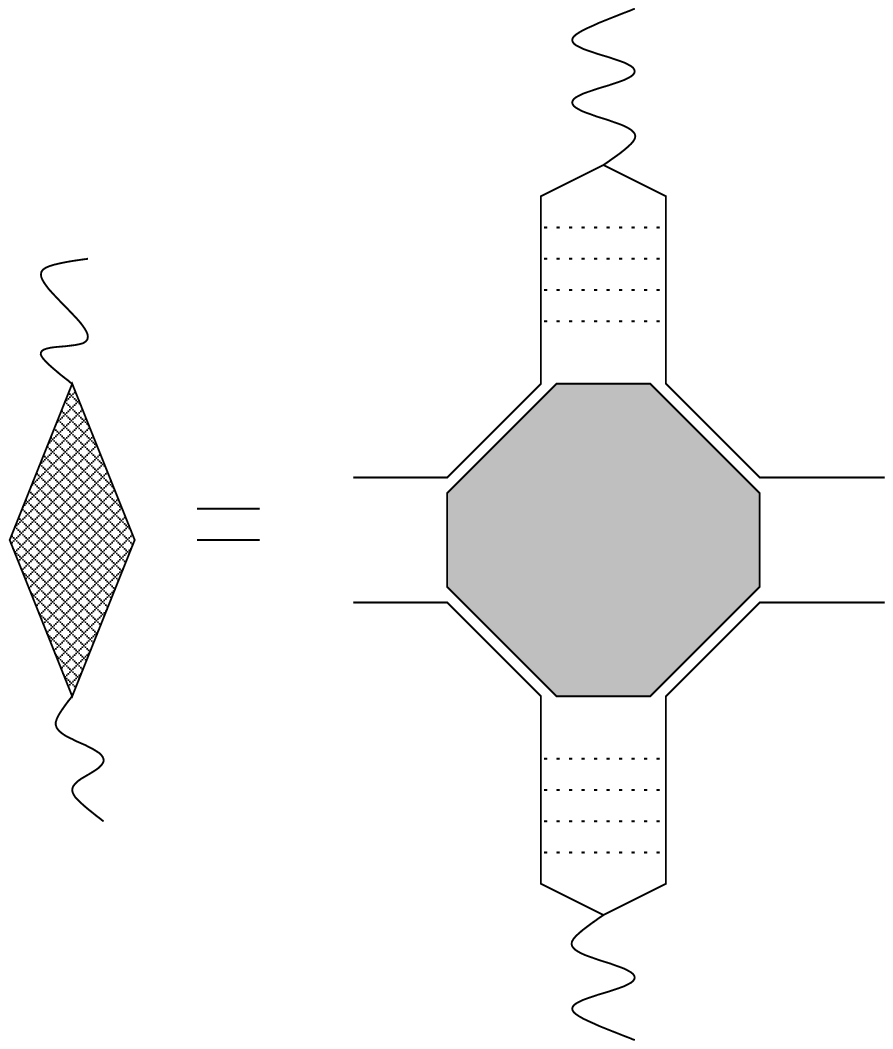}}
\vspace{0.2cm}
\caption{Diffuson contribution to the average of the 
product of two triangular vertices. Cooperon contribution 
is obtained by interchanging of vertices $\protect\rho$ 
and $\protect j$ in one of the triangles (the direction of 
arrows should be changed respectively)}
\label{2}
\end{figure}
}

{
\narrowtext
\begin{figure}[ht]
\vspace{0.2 cm}
\epsfxsize=6 cm
\centerline{\epsfbox{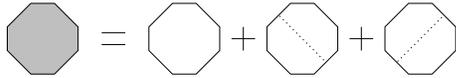}}
\vspace{0.5cm}
\caption{Hikami box.}
\label{3}
\end{figure}
}

In conclusion, we have described the mesoscopic 
fluctuations of the Coulomb drag coefficient or the 
transconductance. The fluctuations are characterized by
the average square of the random, disorder dependent 
transconductance Eq.~(\ref{res}). Compared to the averaged 
transconductance Eq.~(\ref{av}) the fluctuations 
Eq.~(\ref{res}) are determined by the Thouless energy, 
rather than the Fermi energy, as the average. Therefore, 
there exists an intermediate temperature regime, where the 
fluctuations are greater than the average, which results 
in the weak ($1/\sqrt{T}$) temperature dependence of the 
transconductance in this regime. Moreover, in this regime 
the  $\sigma_D$ is a random, sample dependent quantity, 
so that the sign of the measured value is also random. Since 
the average transconductance Eq.~(\ref{av}) grows as 
$T^2$, at higher temperatures ($T>T_*$) the fluctuations 
are small and the the measured $\sigma_D$ is roughly 
equal to the average. This was the case in the existing 
experiments \cite{exp1,exp2,exp3,exp4,exp6}. For the 
samples used in \cite{exp2,exp6} we estimated the 
crossover temperature $T_*\approx0.2 K$, which was below 
the temperature range used in these experiments. To 
observe the effect of the fluctuations, one needs to take 
a dirtier sample of smaller size. Then $T_*$ can be equal 
to several Kelvin, and the saturation of $\sigma_D$ to a 
value with the random sign can be observed.

Finally, we notice that the Coulomb drag coefficient may 
also be presented as the product of two random numbers 
$\rho_D \simeq a_1 a_2$, where $a_1$, $a_2$ characterize 
each layer. If the disorder is corelated, the average 
$\langle a_1 a_2 \rangle$ appears, which leads to the 
results of Ref.~\cite{khve}. In this respect, the results 
of Ref.~\cite{khve} are just a particular manifestation of 
the mesoscopic fluctuations of $\rho_D$, discussed in this 
Letter.

We acknowledge helpful conversations with B.L. Althsuler, 
Fei Zhou, and especially with A. Kamenev. The work at 
Ruhr-Universit\"at Bochum was supported by SFB 237 
``Unordnung and grosse Fluctuationen''. I.A. is A.P. 
Sloan and Packard research fellow.

\end{multicols}

\begin{references}
\bibitem{the1} M.B. Pogrebinskii, Fiz. Tekh. Poluprovodn.  
{\bf 11}, 637 (1977).
\bibitem{the2} P.M. Price, Physica (Amsterdam) 
{\bf 117B}, 750 (1983).
\bibitem{exp1} P.M. Solomon, P.J. Price, D.J. Frank,
and D.C. La Tulipe, Phys. Rev. Lett. 
{\bf 63}, 2508 (1989).
\bibitem{exp2} T.J. Gramila, J.P. Eisenstein, A.H. MacDonald,
L.N. Pfeiffer, and K.W. West, Phys. Rev. Lett. 
{\bf 66}, 1216 (1991).
\bibitem{exp3} U. Sivan, P.M. Solomon, and H. Shtrikman, 
Phys. Rev. Lett. {\bf 68}, 1196 (1992).
\bibitem{exp4} N. Giordano and J.D. Monnier, Phys. Rev. B, 
{\bf 50}, 9363 (1994).
\bibitem{exp6} M.P. Lilly,J.P. Eisenstein,
L.N. Pfeiffer, and K.W. West, Phys. Rev. Lett. 
{\bf 80}, 1714 (1998).
\bibitem{phon} T.J. Gramila, J.P. Eisenstein, A.H. MacDonald,
L.N. Pfeiffer, and K.W. West, Phys. Rev. B, 
{\bf 47}, 12957 (1993).
\bibitem{pho2} H.C. Tso, P. Vasilopoulos, and F.M. 
Peeters, Phys. Rev. Lett. {\bf 68}, 2516 (1992).
\bibitem{oreg} A. Kamenev and Y. Oreg, Phys. Rev. B, 
{\bf 52}, 7516,(1995)
\bibitem{tmp} A.V. Anisovich, B.L. Altshuler, A.G. Aronov, 
and A.Yu. Zyuzin, Pis'ma Zh. 
Eksp. Teor. Fiz. {\bf 45}, 237 (1987) [Sov. Phys. JETP.
Lett. {\bf 45}, 295 (1987)].
\bibitem{pmp} Fei Zhou, B.L. Altshuler, and B.Z. Spivak,
Phys. Rev. Lett. {\bf 82}, 608 (1999).
\bibitem{abr} M. Abramowitz and I. Stegun, 
{\em Handbook of Mathematical Functions} 
(Dover, New York, 1972).
\bibitem{al} I.L Aleiner, B.L. Altshuler, and 
M.E. Gershenson, Waves in Random Media, (1999).
\bibitem{aa} B.L. Altshuler and A.G. Aronov, in 
{\em Electron-Electron Interactions in Disordered 
Systems},eds. A.L. Efros, M. Pollak (North-Holland, 
Amserdam, 1985).
\bibitem{khve} I.V. Gornyi, A.G. Yashenkin, and D.V. 
Khveshchenko, Phys. Rev. Lett. {\bf 83}, 152 (1999).
\end{references}
\end{document}